\documentclass[useAMS,usenatbib]{mn2e}
\usepackage{graphicx}

\def\max{{\rm max}}

\def\sj{{\scriptscriptstyle (j)}}
\def\j{{\scriptscriptstyle (j)}}
\def\tot{{{\rm tot}}}
\def\sub{{{\rm sub}}}
\def\obs{{{\rm obs}}}
\def\dep{{{\rm dep}}}
\def\dur{{{\rm dur}}}
\def\N{\nonumber}
\def\f{\frac}

\def\gtrsim{\ga}
\def\lesssim{\la}

\title[Tail Emission of Prompt GRBs]
{Tail Emission of Prompt Gamma-Ray Burst Jets
}
\author[R. Yamazaki et al.]
{Ryo Yamazaki$^{1,2}$\thanks
{E-mail: ryo@theo.phys.sci.hiroshima-u.ac.jp (RY)},
Kenji Toma$^{3}$,
Kunihito Ioka$^{3}$
and Takashi Nakamura$^{3}$ \\
$^{1}$Department of Physics, Hiroshima University, 
Higashi-Hiroshima, Hiroshima 739-8526,  Japan\\
$^{2}$Department of Earth and Space Science,
Osaka University, Toyonaka 560-0043, Japan\\
$^{3}$Department of Physics, Kyoto University,
Kyoto 606-8502, Japan 
}
\begin{document}

\date{
Accepted 2006 March 08.  
Received 2006 March 08; 
in original form 2005 December 07 
}

\pagerange{\pageref{firstpage}--\pageref{lastpage}} \pubyear{2006}

\maketitle

\label{firstpage}

\begin{abstract}
Tail emission of the prompt gamma-ray burst (GRB) is discussed using 
a multiple emitting sub-shell (inhomogeneous jet, sub-jets or mini-jets) model,
where the whole GRB jet consists of many emitting sub-shells.
One may expect that such a jet with angular inhomogeneity should
produce spiky tail emission.
 However, we found that the tail is not spiky but is
 decaying roughly  monotonically.
The global decay slope  of the tail is not so much affected by
the local angular inhomogeneity but affected by
the global sub-shell energy distribution.
The fact that steepening GRB tail breaks appeared in some events 
prefers the structured jets.
If the angular size of the emitting
sub-shell is around 0.01--0.02~rad, 
some bumps or fluctuations appear in the tail emission
observed frequently in long GRBs.
If the parameter differences of sub-shell properties are large, the 
tail has frequent changes of the temporal slope observed
in a few bursts.
Therefore, the multiple emitting sub-shell model has
 the advantage of explaining the small-scale structure
 in the observed rapid decay phase.

\end{abstract}


\section{Introduction}
\label{sec:intro}

Gamma-ray bursts (GRBs) consist of two phases:
prompt GRB emission and subsequent afterglows.
How long the prompt GRB emission
lasts and when the transition from the prompt GRB to the afterglow occurs
have been long-standing problems.
These problems are tightly related to the mechanism of the central
engine of GRBs.
In general, the prompt GRB tends to show a spectral softening and 
a rapid decay \citep[e.g.,][]{giblin99,connaughton02}, so that
the X-ray observation with high flux sensitivity is necessary
to investigate the end epoch of the prompt GRB.
Such an observation has become possible
thanks to the {\it Swift} satellite, which are revealing rich structures
in early X-ray counterparts of GRBs
\citep{burrows05,chincarini05,campana05,tagliaferri05,nousek05,obrien06}.
A distinct decaying component before the usual afterglow phase
is identified.
During this epoch, 
the decay is steeper and the spectral index is different
compared with the subsequent phase \citep{nousek05}.
These results suggest that this component is 
the tail emission of the prompt GRB \citep{zhang05,pana05}.

Most natural explanation for the tail is a high latitude emission
from a relativistically moving shell \citep{kumar00ng,yama05}.
Suppose the shell shines for a short period. 
Since the shell has a curvature, photons far from the line of sight come later. 
Because the shell at higher latitude 
from the line of sight has a lower velocity toward the observer,  
the emission becomes dimmer and softer as time passes 
because of the relativistic beaming effect.
For a spherical uniform shell, the predicted decay index, $\alpha$,
is related to the photon index, $\beta(<0)$, as
$\alpha=-1+\beta$ \citep{kumar00ng}.

However, there are many difficulties when we interpret the
observed prompt GRB tails with the current model
in which the uniform, instantaneous emission is assumed \citep{kumar00ng}.
The observed $\alpha$--$\beta$ relation for the steep decay phase does
not tightly obey the relation, $\alpha=-1+\beta$, but has a wide
scatter. This suggests that  several additional factors are needed
to spread the relation.
Some events showed the steepening break in the tail \citep{obrien06}, 
which we call ``the steepening GRB tail break'' in order to discriminate it
from other kinds of breaks appearing in the afterglow.
This fact  suggests that the assumption of the uniform shell
in the current model should be modified.
Moreover,
some light curves in the rapid decay phase are neither
simply monotonic nor smooth, but show some bumps and dips
\citep[e.g., see Fig.~1 of][]{tagliaferri05}.
Some bursts  showed frequent changes in the temporal slope of the rapid decay
phase \citep{obrien06}.
These small-scale features are not produced by the smooth jet
with angular homogeneity.
Hence we should modify the current model in order to explain
these detailed observed features.
Since the emission region sweeps the shell spatially,
the tail emission features, e.g., the decay index and smoothness,
would diagnose the unknown angular structure of the GRB jet.

In this paper, we study the tail emission of the prompt GRB
 using a multiple emitting ``sub-shell'' 
(inhomogeneous jet, sub-jets or mini-jets) model considered so far
 \citep{nakamura00,kumar00,yama04b,toma05,toma05b}.
Standard GRB scenario assumes the  density fluctuation
in the radial direction,
and the high-density region is called the ``shell''.
Although the matter exists outside of the shell in general, 
we usually adopt the null-density approximation there.
Similar to the radial direction,
one can expect that the GRB outflow has the inhomogeneity also
in the angular direction, because causally connected region has
an angular size of $\gamma^{-1}$, which is an order of magnitude 
smaller than that of the GRB jet inferred from the afterglow observations.
We further develop this picture and consider the emission region is 
patchy. We call each patch as sub-shell. Similar to null-density
approximation in the shell, we assume that
 outside of the sub-shell, there is little emission.
Therefore, multiple sub-shell approximation is along a natural extension
of current standard scenario, though it is an extreme model 
to express  the local angular inhomogeneity. One may consider our
sub-shell model as a finite difference version of the continuous 
variable in numerical simulations. 
Our basic claim is that the relativistic
kinematics and the viewing angle effect are one of the most important
parameters that causes observed properties of GRBs. At present it is
important to investigate pure kinematical effects with other
parameters fixed.
Our previous works have shown that the
multiple sub-shell model can explain the observed diversity of prompt
GRB emission. 
{\it Swift} observations show the diversity of early X-ray afterglow light curves,
which might be ascribed to the angular inhomogeneity of GRB jets.
So far, no one has investigated the prompt GRB tail using the
multiple sub-shell model, though the
subsequent afterglow phase arising from external shocks with
angular inhomogeneity has been calculated
\citep[e.g.,][]{rossi02}.
This paper is organized as follows.
In \S~\ref{sec:model}, we briefly introduce our prompt emission model.
The tail emission of GRB is discussed in \S~\ref{sec:tail}.
In \S~\ref{sec:obs}, we will see that our model well reproduces 
the observed features of the tail emission of the prompt GRB.
Section~\ref{sec:dis} is devoted to discussions.

\section{Prompt emission model}
\label{sec:model}

We consider the same model as discussed in our previous works
\citep{yama04b,toma05,toma05b}.
The whole GRB jet, whose opening half-angle is $\Delta\theta_{\tot}$,
consists of $N_{\tot}$ emitting sub-shells (sub-jets)
\citep{nakamura00,kumar00}.
We introduce the spherical coordinate system $(r, \vartheta, \varphi, t)$ in
the central engine frame, where the origin is at the
central engine, and $\vartheta=0$ is the axis of the whole jet.
Each sub-jet departs at time $t_{\dep}^{\sj}$ 
($0<t_{\dep}^{\sj}<t_{\rm dur}$, where $j=1,\cdots, N_{\tot}$, and 
$t_{\rm dur}$ is the active time of the central engine) 
from the central engine in the direction of 
$\vec{n}^{\sj}=(\vartheta^{\sj}, \varphi^{\sj})$.
The direction of the observer is denoted by 
$\vec{n}_{\rm obs}=(\vartheta_{\obs}, \varphi_{\obs})$.
For each sub-jet, the emission model is the same as in the previous works
\citep{granot99,woods99,ioka01,yama02,yama03,yama04a}.
The observed flux from the $j$th sub-jet
is calculated when the following parameters are determined:
the viewing angle of the sub-jet $\theta_v^{\sj}
=\cos^{-1}(\vec{n}_{\rm obs}\cdot\vec{n}^{\sj})$,
the opening half-angle of the sub-jet $\Delta\theta_{\sub}^{\sj}$,
the departure time $t_{\dep}^{\sj}$,
the Lorentz factor $\gamma^{\sj}=(1-\beta_\sj^2)^{-1/2}$, 
the emitting radius $r_0^{\sj}$, 
the low- and high-energy photon index $\alpha_B^{\sj}$ and $\beta_B^{\sj}$, 
the break frequency in the shell comoving frame ${\nu'_0}^{\sj}$ \citep{band93},
the normalization constant of the emissivity $A^{\sj}$, 
and the source redshift $z$.
The whole light curve from the GRB jet is produced by the superposition
of the sub-jet emission.
Throughout the paper, we neglect the cosmological effect,
i.e., we set $z=0$.

As we will see later, the local inhomogeneity in our model is almost
averaged during the tail emission phase, and the global jet structure (the mean
sub-jet distribution) determines overall shape of the tail.
Therefore, essentially we are also studying the tail emission from
the usual structured jets at the same time as the limiting case, i.e., 
from uniform or power-law jets
with no local inhomogeneity.

\section{Tail Emission of Prompt GRB}
\label{sec:tail}

\subsection{General Kinematical Considerations}
\label{subsec:kinematic}

We present some of kinematical properties of prompt GRBs in
the multiple sub-jet model.
Let $\theta_v^\sj$ be the angle between the
observer's line of sight and the axis of the $j$th sub-jet.
The pulse starting and ending time at the observer are given by
\begin{eqnarray}
T_{\rm start}^\sj &\sim&
t_{\dep}^\sj+
\f{r_0^\sj}{2c\gamma_\sj^2}\left(1+\gamma_\sj^2{\theta_-^\j}^2
\right)~~ ,
\label{eq:Tstart}\\
T_{\rm end}^\j &\sim& 
t_\dep^\j+
\f{r_0^\j}{2c\gamma_\j^2}\left(1+\gamma_\sj^2{\theta_+^\j}^2
\right) ~~,
\label{eq:Tend}
\end{eqnarray}
where $\theta_+^\j=\theta_v^\j+\Delta\theta_\sub^\j$ and
$\theta_-^\j=\max\{ 0,\theta_v^\j-\Delta\theta_\sub^\j \}$, and
we use the formulae $\beta_\sj\sim1-1/2\gamma_\sj^2$
and $\cos\theta\sim1-\theta^2/2$ 
for $\gamma^\sj\gg1$ and $\theta\ll1$, respectively.
The observer time $T=0$ is chosen as the time of arrival at the 
observer of a photon emitted at the origin $r=0$ at $t=0$.
Then, the pulse duration, $\delta T^\sj=T_{\rm end}^\sj-T_{\rm start}^\sj$,
is given by
$\delta T^\sj\sim1.5~r_{14}(\theta_+^\j/0.03)^2$~s
for $\theta_v^\sj<\Delta\theta_\sub^\sj$, and
\begin{equation}
\delta T^\sj\sim26~r_{14}
(\Delta\theta_\sub^\j/0.02)(\theta_v^\j/0.2)~{\rm sec}~~, 
\label{eq:deltaTsub}
\end{equation}
for $\theta_v^\sj>\Delta\theta_\sub^\sj$,
where $r_{14}=r_0^\j/10^{14}~{\rm cm}$.
The peak energy $E_{\rm p}^\j$, that gives the peak 
of the $\nu F_\nu$ spectrum, is approximated as
$E_{\rm p}^\j\propto{\nu'_0}^\j\delta^\j$, where
$\delta^\j=[\gamma^\j(1-\beta_\j\cos\theta_-^\j)]^{-1}$.
Using practical numerical calculations, we find
$E_{\rm p}^\j\sim6.5\times10^2\gamma_2\nu_{5}$~keV
for $\theta_v^\sj\ll\Delta\theta_\sub^\sj$, and
\begin{equation}
E_{\rm p}^\j\sim2~{\gamma_2}^{-1}\nu_{5}
(\theta_v^\j/0.2)^{-2}~{\rm keV}~~,
\label{eq:Ep}
\end{equation}
for $\theta_v^\sj\gg\Delta\theta_\sub^\sj$,
where $\nu_{5}={\nu'_0}^\j/5~{\rm keV}$ and $\gamma_2
=\gamma^\j/10^2$ \citep[see also][]{graziani06}.
We note that the light curve from a single pulse becomes dim
and smooth for large $\theta_v^\j$ \citep[see Fig.~2 of][]{ioka01}.

For sub-jets with $\theta_v^\j \lesssim \Delta\theta_\sub^\j$,
we have $\delta T^\sj \ll t_\dur$, because $t_\dur$ is about several
tens of seconds.
Then the bright pulses arising from these sub-jets are concentrated 
in the epoch $0 \lesssim T \lesssim t_\dur$.
On the other hand, for sub-jets with $\theta_v^\j \gtrsim 0.1~r_{14}^{-1/2}
(t_\dep^\j/20~\rm{sec})^{1/2}$, the second terms of the right hand sides of 
Eqs.~(\ref{eq:Tstart}) and (\ref{eq:Tend}) become larger than
the first ones, so that $T_{\rm start}^\j$ and $T_{\rm end}^\j$
may be  much larger than $t_\dur$ for large $\theta_v^\j$.
Such smooth, long-duration, dim and soft pulses make the tail emission 
of the prompt GRB.

Now, we concentrate on the temporal structure of the tail emission.
We assume that each sub-jet is not drastically different.
Since $T_{\rm start}^\j$ and $T_{\rm end}^\j$
are much larger than $t_\dur$, they are not affected by $t_\dep^\j$. 
Thus we may consider that all the sub-jets contributing to the tail emission
emit simultaneously at a mean radius $r_0$ in the central engine frame.
The end of the tail emission $T_{\rm tail}$ is determined by the angular 
size of the whole jet:
\begin{eqnarray}
T_{\rm tail} &\sim& (r_0/2c)(\Delta\theta_\tot + \vartheta_\obs)^2 \N\\
&\sim& 2 \times 10^2 r_{14} [(\Delta\theta_\tot + \vartheta_\obs)/0.3]^2
~{\rm sec}~~.
\label{eq:Ttail}
\end{eqnarray}
The tail flux at an observer time $T$ is the superposition of the 
sub-jet emission with the pulse starting and ending time 
$T_{\rm start}^\j < T < T_{\rm end}^\j$.
Sub-jets with viewing angles between
$\theta_T-\Delta\theta_\sub$ and $\theta_T+\Delta\theta_\sub$
contribute to the tail flux at a time $T$, where
$\theta_T=(2cT/r_0)^{1/2}\sim0.2r_{14}^{-1/2}(T/10^2{\rm s})^{1/2}$~rad.
Then, we may calculate the number of these contributing sub-jets
$N_\sub(T)$ and  its variance $1/\sqrt{N_\sub(T)}$ 
if the sub-jet distribution and observer's line of sight are given.
Since $N_\sub(T)$ is sufficiently large, the tail light curve will be smooth.
In the following, we actually compute the light curves of prompt GRB
emission for various cases.

\subsection{Example 1. Uniform jet}
\label{subsec:uniform}

We first consider the uniformly distributed sub-jets.
The number of sub-jets per unit solid angle is approximately given by
$dN/d\Omega=N_\tot/(\pi\Delta\theta_\tot^2)$
for $\vartheta<\Delta\theta_\tot$,
where $\Delta\theta_\tot=0.25$~rad is adopted.
The departure time of each sub-jet $t_{\dep}^{\sj}$ is assumed to be
homogeneously random between $t=0$ and $t=t_{\dur}=20$~sec.
The central engine is assumed to produce $N_{\tot}=1000$ sub-jets.
In this section, we assume that all the sub-jets have the same 
values of the following fiducial parameters:
$\Delta\theta_{\sub} = 0.02$~rad, $\gamma=100$,
$r_0 = 1.0 \times 10^{14}$~cm, $\alpha_B = -1$,
$\beta_B = -2.3$, $h{\nu'_0}=5$~keV,
and $A={\rm const}$.

\begin{figure}
\includegraphics[width=84mm]{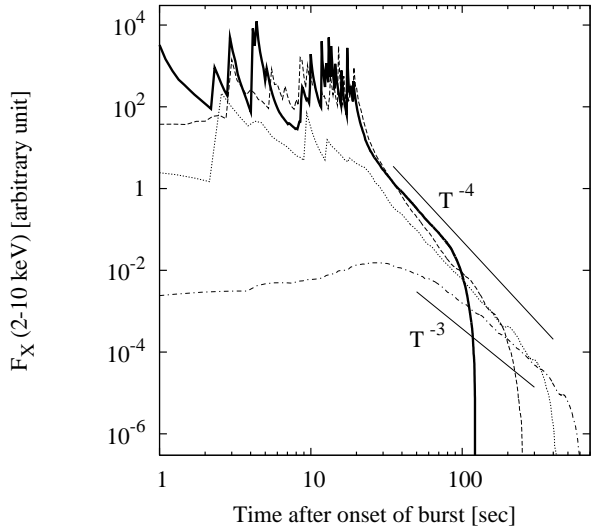}
 \vspace{0pt}
 \caption{
Examples of light curves of the prompt GRB emission
calculated by the multiple emitting sub-shell model 
(multiple sub-jet model).
The sub-jets are distributed uniformly.
All sub-jets have the same intrinsic properties.
Thick-solid, dashed, dotted and dot-dashed lines correspond to
the viewing angles of $\vartheta_\obs=0$,  $\Delta\theta_\tot/2$, 
$\Delta\theta_\tot$ and $3\Delta\theta_\tot/2$, respectively.
The observer time is the time since the onset of the burst.
}
\label{fig1}
\end{figure}

Figure~\ref{fig1} shows the results, where 
$\vartheta_\obs=0$ (thick-solid line), $\Delta\theta_\tot/2$ (dashed), 
$\Delta\theta_\tot$ (dotted), and 
$3\Delta\theta_\tot/2$ (dot-dashed) are considered.
One can see the entire behavior of  bursts.
For the cases $\vartheta_\obs<\Delta\theta_\tot$, 
as expected, the bright pulses are observed in
the period $0\lesssim T\lesssim t_\dur=20$~sec, and
subsequently the tail emission starts.
However, when $\vartheta_\obs=3\Delta\theta_\tot/2$,
the whole jet is seen off-axis. Then the emission becomes dim and soft 
\citep{yama02,yama03,yama04a,yama04b}.
In any cases, the tail emission is smooth. 
For $\vartheta_\obs=0$, the number of contributing sub-jets
$N_\sub(T)$ is approximately given by 
$N_\sub(T)\sim4\pi\theta_T\Delta\theta_\sub(dN/d\Omega)$.
For our adopted parameters, we derive
$N_\sub(T)\sim2.6\times10^2~(T/10^2{\rm s})^{1/2}\gg1$,
so that the tail light curve is smooth.
Also in other cases ($\vartheta_\obs\ne0$), we obtain 
$N_\sub(T) \gg 1$.

\begin{figure}
\includegraphics[width=84mm]{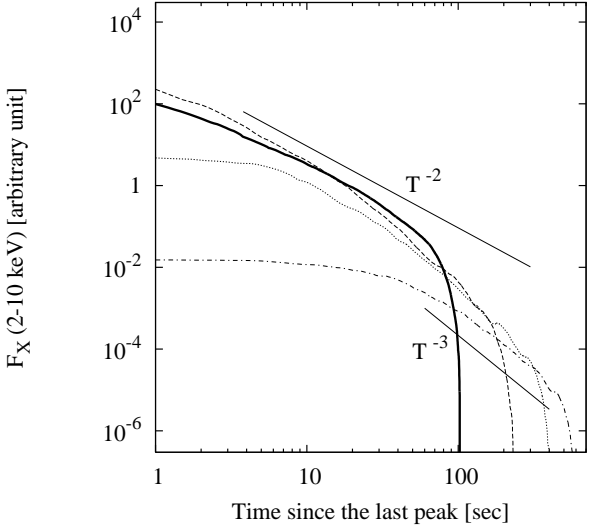}
 \vspace{0pt}
 \caption{
The same as Fig.~\ref{fig1}, but
the observer time is the time since the last bright peak.
}
\label{fig1b}
\end{figure}

The decay index of the tail emission, $\alpha$, 
is about $-4$ when it is determined by
the whole light curve (see Fig~\ref{fig1}).
The effect of choosing the zero of time is essential in order to
determine $\alpha$ \citep{chincarini05,zhang05}.
Figure~\ref{fig1b} shows the same light curves as in Fig.~\ref{fig1}
but the time zero is shifted to the maximum of the last bright pulse.
Then, we find $-3\lesssim\alpha\lesssim-2$.
Since we may consider that all the sub-jets emit simultaneously in the central 
engine frame for the tail emission, 
the behavior of the emission may be similar for a single, spherical,
infinitesimally thin shell case.
Then  $\alpha=-1+\beta_B=-3.3$ is expected \citep{kumar00ng},
which is intermediate between those shown in Figures~\ref{fig1}
and \ref{fig1b}.
This implies that the zero of time which gives us 
$\alpha=-1+\beta_B$ lies between the onset of the burst and the last
bright pulse.
Therefore, it is difficult to check whether the relation
$\alpha=-1+\beta_B$ is satisfied or not.

\subsection{Example 2. Power-law jet}
\label{subsec:powerlaw}

Next, we  consider the power-law sub-jet distribution, i.e.,
$dN/d\Omega=C$
for $0<\vartheta<\vartheta_c$ and
$dN/d\Omega=C(\vartheta/\vartheta_c)^{-2}$ for
$\vartheta_c<\vartheta<\Delta\theta_\tot$,
where $\vartheta_c=0.03$~rad and $\Delta\theta_\tot=0.3$~rad,
and $C=(N_\tot/\pi\vartheta_c^2)[1+2\ln(\Delta\theta_\tot/\vartheta_c)]^{-1}$
is the normalization constant
\citep{rossi02,zhang02,toma05,toma05b}.
The departure time of each sub-jet $t_{\dep}^{\sj}$ is assumed to be
homogeneously random between $t=0$ and $t=t_{\dur}=20$~sec and we adopt
$N_{\tot}=350$.
We assume that all the sub-jets have the fiducial parameters.

\begin{figure}
\includegraphics[width=84mm]{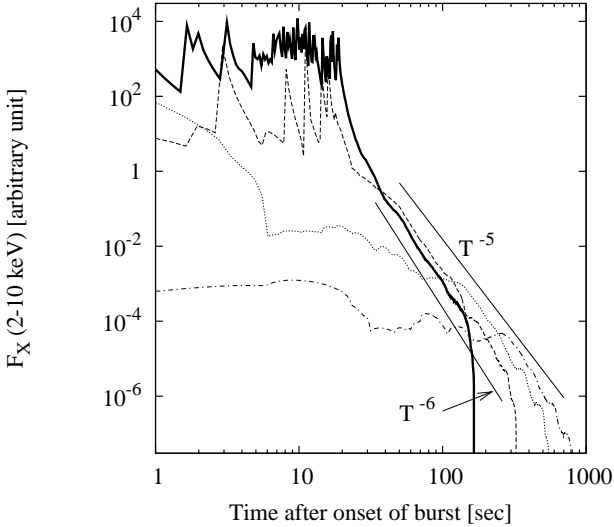}
 \vspace{0pt}
 \caption{
The same as Fig.~\ref{fig1} but for the
power-law sub-jet distribution.
}
\label{fig2}
\end{figure}

\begin{figure}
\includegraphics[width=84mm]{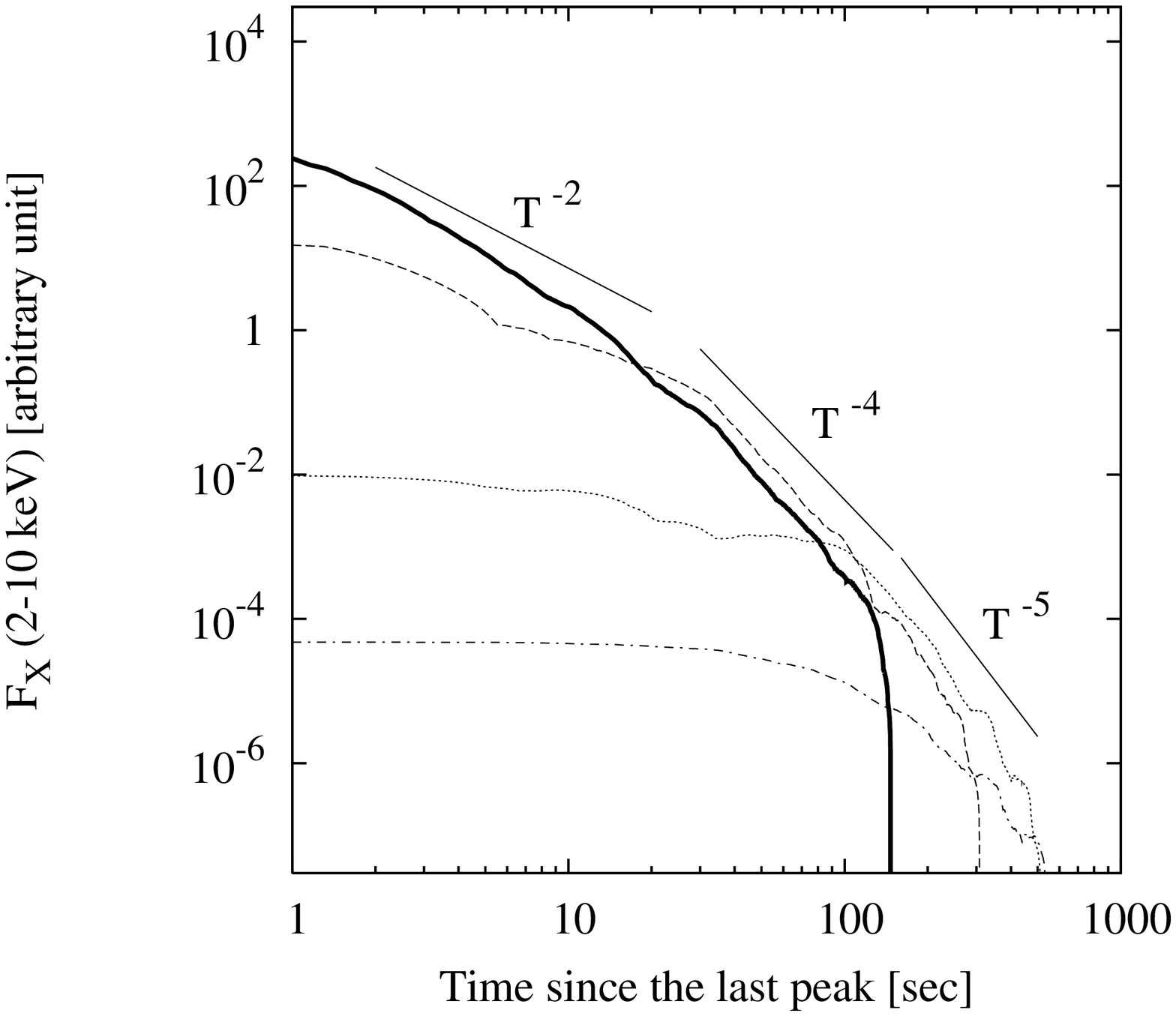}
 \vspace{0pt}
 \caption{
The same as Fig.~\ref{fig1b} but for the
power-law sub-jet distribution.
}
\label{fig2b}
\end{figure}

Figures~\ref{fig2} and \ref{fig2b} show the results.
Compared with the uniform jet case, the decay  is steeper
because the power-law jet is dimmer in the outer region,
i.e., the sub-jets are sparsely distributed near the periphery of
the whole jet.
For example, when $\vartheta_\obs=0$, the number of contributing sub-jets
$N_\sub(T)$ is calculated as
$N_\sub(T)\sim4\pi\theta_T\Delta\theta_\sub(dN/d\Omega)
\sim28~(T/10^2{\rm s})^{-1/2}$, which is
a decreasing function of time contrary to
the uniform jet case.

\section{Comparison between theoretical and observational results}
\label{sec:obs}

As can be seen in Fig.~\ref{fig2b}, a steepening GRB tail break 
(from $\alpha\sim-1$ to $\alpha\sim -4$) occurs
at about 35~sec and 100~sec after the last brightest pulse
for $\vartheta_\obs=\Delta\theta_\tot/2$ and
$\vartheta_\obs=\Delta\theta_\tot$, respectively.
This is because the power-law jet has a core region
($0<\vartheta<\vartheta_c$), where sub-jets densely distributed
compared with the outer region.
Before photons emitted by the core arrive at the observer
(i.e., $\theta_T\lesssim\vartheta_\obs$),
$N_\sub(T)$ increases with $T$ more rapidly than in the case
of the uniform sub-jet distribution.
Then, the light curve shows a shallow decay 
or even shows a rising part.
After the photons arising from the core are observed,
the sub-jet emission with viewing angles larger than $\sim\vartheta_c$
is observed. Then $N_\sub(T)$ rapidly decreases with $T$ and
the observed flux suddenly drops. Therefore,
a steepening GRB tail break occurs at $\theta_T\sim\vartheta_\obs$.
{\it Swift} observations have shown that some 
bursts (e.g. GRB 050421, 050713B) have the
steepening GRB tail break, which may support the structured jet case.

\begin{figure}
\includegraphics[width=84mm]{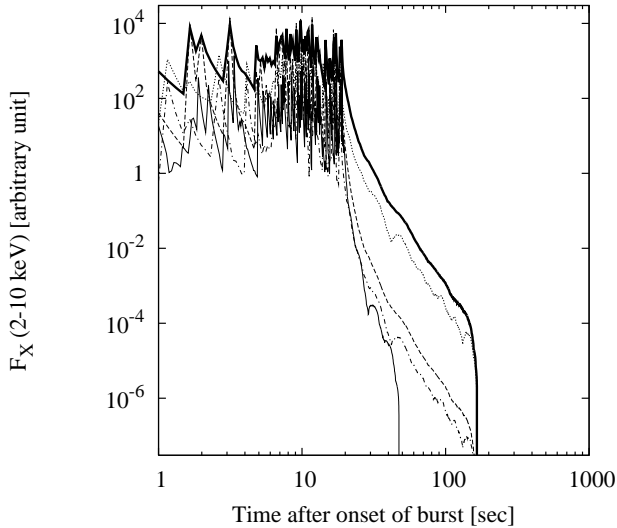}
 \vspace{0pt}
 \caption{
Comparison of light curves with different parameter sets of
$\Delta\theta_\sub$, $\gamma$ and $r_0$ for the 
power-law distribution.
Dashed ($\gamma=400$), dotted
($\Delta\theta_\sub=0.01$~rad), 
dot-dashed ($\gamma=400$ and $\Delta\theta_\sub=0.01$~rad), 
and thin-solid lines ($r_0=0.2\times10^{14}$~cm)
with the other parameters being fiducial, i.e., 
$\Delta\theta_\sub=0.02$~rad, $\gamma=100$ and 
$r_0=1.0\times10^{14}$~cm.
The thick-solid line is the same as that in
Fig.~\ref{fig2} (fiducial set).
All lines have $\vartheta_\obs=0$.
}
\label{fig3}
\end{figure}

Observed  rapid decay phase lasts
at least $\sim100$~sec in the cosmological rest frame \citep{nousek05},
which constrains the emitting radius $r_0$.
As can be seen in Fig.~\ref{fig3}, 
when $r_0$ is small (the thin-solid line)
the tail emission ends rapidly,
while the end time of the tail emission, 
$T_{\rm tail}$, does not depend on $\Delta\theta_\sub$ and 
$\gamma$ [see Eq.~(\ref{eq:Ttail})]. 
The end time of the tail is difficult to be determined because 
the subsequent shallow decay arising from the external shock 
overlies the prompt tail emission. 
So, we can set $T_{\rm tail}\gtrsim10^2$~sec, and obtain 
$r_0\gtrsim5\times10^{13}[(\Delta\theta_\tot 
+ \vartheta_\obs)/0.3]^{-2}$cm.

Observed light curves in a rapid decay phase are neither
simply monotonic nor smooth, but show some bumps and dips
\citep[e.g., see Fig.~1 of][]{tagliaferri05}.
These small-scale features are not explained by the angularly 
continuous jet model. We claim that they
 come from the inhomogeneity of the emission region.
In calculating the light curves shown in Figs.~1--4,
we assumed $\Delta\theta_\sub=0.02$~rad, so that the local
inhomogeneity was erased and smooth monotonic decay behavior
appeared.
If $\Delta\theta_\sub$ becomes small, we can obtain the light curve 
with small bumps similar to the observed one.
Figure~\ref{fig3} shows the dependence on the assumed parameters
on $\Delta\theta_\sub$ and $\gamma$ (but still the same for all $j$),
where we fix the observer's line of sight $\vartheta_\obs=0$
and the power-law sub-jet distribution is considered.
When $\Delta\theta_\sub$ is small (dotted and dot-dashed lines), 
$\delta T^{\j}$ is small [see Eq.~(\ref{eq:deltaTsub})],
so that $N_\sub(T)$ decreases.
Then the observed light curves
remarkably reflect fluctuations of $N_\sub(T)$.
On the other hand,
when  $\gamma$ is large  but other parameters are
fixed (dashed and dot-dashed lines),
the smoothness of the light curve remains unchanged
because $N_\sub(T)$ does not depend on $\gamma$.
Then, the asymptotic decay slope is unchanged, though
the flux becomes dim for large $\gamma$ because 
the relativistic beaming effect is large.

\begin{figure}
\includegraphics[width=84mm]{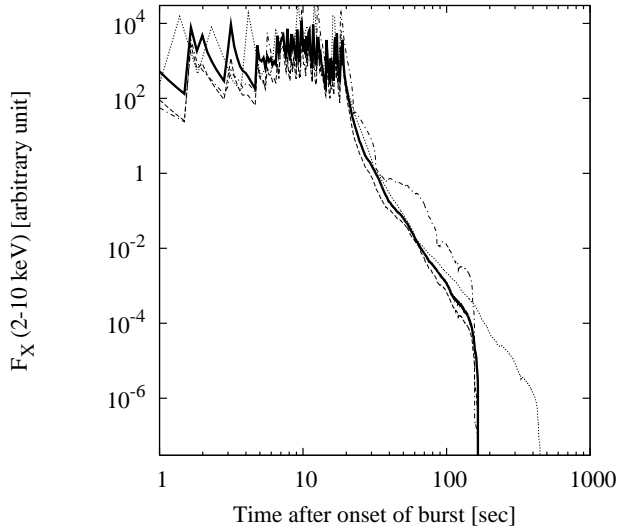}
 \vspace{0pt}
 \caption{
Comparison of light curves with different parameter sets of
$\Delta\theta_\sub$, $\gamma$ and $r_0$ for the 
power-law distribution.
The value of $\gamma{\nu'_0}^\j$ is distributed randomly according to 
the log-normal distribution with an average of $\log(350~{\rm keV})$
and a logarithmic variance of 0.2, and $A^\j$ is determined so that
the observed flux is proportional to $\xi E_p^2$ for $\theta_v^\j=0$, 
where $\xi$ is also assumed to obey a log-normal distribution
with a logarithmic variance of 0.15 \citep{lloyd02,yone04,liang04}.
The dashed line corresponds to the fiducial parameter set of
$\Delta\theta_\sub$, $\gamma$ and $r_0$, while
the dotted line is for $\gamma=300$ and $r_0=3.0\times10^{14}$~cm.
The dot-dashed line is for the case in which logarithmic variance of
$\gamma{\nu'_0}^\j$ is changed into 0.8.
The thick-solid line is the same as that in
Fig.~\ref{fig2} (fiducial set).
All lines have $\vartheta_\obs=0$.
}
\label{fig3b}
\end{figure}

Some bursts  showed frequent changes in a temporal slope of the rapid decay
phase \citep{obrien06}.
For example, GRB~050819 exhibited  steep decay with small fluctuations
until $\sim250$~sec after the BAT trigger. Subsequently,
the  decay became shallow and lasted for $\sim250$~sec, and the rapidly
decaying tail started again at $\sim500$~sec after the BAT trigger.
We find that when the differences of  sub-jet properties are small,
such observed behavior does not appear.
In Fig.~\ref{fig3b},
the value of $\gamma{\nu'_0}^\j$ is distributed randomly according to 
the log-normal distribution with an average of $\log(350~{\rm keV})$
and a logarithmic variance of 0.2, and $A^\j$ is determined so that
the observed flux is proportional to $\xi E_p^2$ for $\theta_v^\j=0$, 
where $\xi$ is also assumed to obey a log-normal distribution
with a logarithmic variance of 0.15 \citep{lloyd02,yone04,liang04}.
When the variance is small (dashed and dotted lines), 
the change of the slope is small, while
 the shape is similar to that of GRB~050819
for larger variance case (dot-dashed line).
However, the small number of events showing such complicated decay behavior
like GRB~050819 may suggest that the differences of sub-jet properties
are small.

\section{Discussion}
\label{sec:dis}

We have examined the tail emission of the prompt GRB  
in the X-ray band using 
a multiple emitting sub-shell (inhomogeneous jet, sub-jets or mini-jets) model,
and  have confirmed that our inhomogeneous jet model
well reproduces the observed features  of the tail emission.
The sub-shell emission with a large viewing angle causes a smooth,
long-duration, dim, soft pulse and arrives later than the bright hard
spikes. These components make the tail emission of the prompt GRB
similar to the observed one.
Since the pulse duration is long for a large viewing angle,
the local inhomogeneity is almost averaged and the  continuous
light curve is obtained.
The global sub-jet angular distribution determines the shape of the
 global decay slope of the tail.
Therefore, the discrete multiple sub-jet model and
the continuous surface model obtained by averaging sub-jets
predict the same ``averaged'' decay index.
We have found that the decay is steeper for the power-law jet case
than for the uniform jet case.
It has also been found that if there is a core, in which many emitting shells
exist compared with other regions, the steepening GRB tail break appears 
 when the core is viewed off-axis.
Such breaks have been seen for some bursts.
Finally, it has been found that
in the sub-jet model with  $\Delta\theta_\sub\sim0.01$~rad,
the local inhomogeneity is not entirely erased in the tail
and makes small bumps or fluctuations on the smooth decay that
have been observed in many events.

{\it Swift} observation has revealed 
the short-duration and large-amplitude flares
in some X-ray afterglows \citep{burrows05}.
Such X-ray flares during the afterglow phase are 
 thought to be produced by the long-acting 
engine \citep{ioka05}.
In our multiple sub-shell model,
when a sub-shell causes very bright emission with large viewing angle,
the flare-like structure is observed in the rapid decay phase.
Hence the X-ray flare in the rapid decay phase
may be originated in either the long-acting engine or
 the angular inhomogeneity of the emitting jet.

For uniform, instantaneous emission, the decay index, $\alpha$, of
the prompt GRB tail is related to the photon index, $\beta$,
as $\alpha=-1+\beta$ \citep{kumar00ng}. 
As seen in this paper, there are several effects that make
the decay index different from this simple relation:
choosing the time zero, the energy distribution of the jet,
and the viewing angle of the jet.
It is difficult to separate these factors from the observation.
Therefore, the observed $\alpha$--$\beta$ relation 
would have a wide scatter.

A typical value of the viewing angle at an observer time $T$
is $\theta_T\sim0.2(T/10^2{\rm s})^{1/2}$~rad,
and then the peak energy, $E_{\rm p}$, is about a few keV
[see Eq.~(\ref{eq:Ep})].
Therefore, the spectral index in the X-ray band is given by the
high-energy photon index $\beta_B \sim -2.3$, which is consistent with
the observed photon index  that 
ranges between 1.34 and 3.25 (average value is 2.28) \citep{nousek05}.

Important parameters that characterize our model are
the number of sub-jets $N_{\rm tot}$ and the opening half-angle
of the sub-jet $\Delta\theta_\sub$.
Given the sub-jet angular distribution, 
the value of $N_{\rm tot}$ is determined in order to
reproduce the number of sub-jets along a line of sight, 
$n_s$, that ranges between about $10$ and $10^2$ because
$n_s$ corresponds to the number of bright pulses observed in
usual long GRBs \citep{yama04b}.
In our adopted parameters, the maximum of $n_s$ is about 10 and 30
for uniform jet case and the power-law jet case, respectively.
On the other hand, the value of $\Delta\theta_\sub$ has been
 fairly uncertain.
It is a common sense that $\Delta\theta_\sub$ is larger than
$\gamma^{-1}$ because even if 
$\Delta\theta_\sub\ll\gamma^{-1}$ initially,
jet expands sideways and the asymptotic value of $\Delta\theta_\sub$
is $\gamma^{-1}$. 
However, in principle, $\Delta\theta_\sub$ 
could be even smaller than $\gamma^{-1}$
e.g., if the offset collision of two shells is considered. 
In this paper, we have found that the most preferable is the case of
$\Delta\theta_\sub\sim\gamma^{-1}\sim0.01$~rad.
If  $\Delta\theta_\sub\ll0.01$, the tail would be so spiky,
while if  $\Delta\theta_\sub\gg0.01$, the tail would be so smooth 
that small-scale bumps would disappear.

The behavior of afterglow light curves of GRBs may  also be
a diagnostic tool to investigate the jet structure 
\citep{rossi02,granot03,pana03}.
However,  the prompt GRB tail is better 
 than the afterglow light curve for this purpose.
This is because the information of the jet
structure is lost at later time due to the hydrodynamical energy
re-distribution effect \citep{granot03} and the
tail emission reflects the GRB jet structure more directly.

\section*{Acknowledgments}

We would like to thank the anonymous referee,
T.~Sakamoto, G.~Sato, T.~Takahashi 
and D.~Yonetoku for useful comments and discussions.
This work was supported in part by
 a Grant-in-Aid for the 21st Century COE
 ``Center for Diversity and Universality in Physics''
 and also by 
Grants-in-Aid for Scientific Research of the Japanese Ministry 
of Education, Culture, Sports, Science and Technology 
09245 (R.~Y.), 14047212 (T.~N., K.~I.), 
14204024 (T.~N.) and 17340075 (T.~N.).

%

%\appendix
%\section[]{}

%\bsp

\label{lastpage}

\end{document}